\documentstyle[12pt,epsf]{article}

\begin{document}
\renewcommand{\dag}{{\scriptscriptstyle{+}}}

\newcommand{\sheptitle}
{Neutrino Masses and Mixing Angles in a Supersymmetric
SU(4)$\otimes$SU(2)$_L\otimes$SU(2)$_R$ Model}

\newcommand{\shepauthor}
{B. C. Allanach\footnote{Address after 1 October 1995: Rutherford
Appleton Laboratory, Chilton, Didcot, OX11 0QX.} and S. F. King}

\newcommand{\shepaddress}
{Physics Department, University of Southampton\\Southampton, SO17 5NH,
U.K.}

\newcommand{\shepabstract}
{We consider the problem of neutrino masses and mixing angles in a
supersymmetric model based on the gauge group
SU(4)$\otimes$SU(2)$_L\otimes$SU(2)$_R$ broken at the scale
$M_X\approx 10^{16}$ GeV.  We extend a previous operator analysis of
the charged lepton and quark masses and mixing angles in this model to
include the neutrino sector, assuming a universal Majorana mass $M$
for the right-handed neutrinos. The Dirac part of the neutrino matrix
is then fixed and the physical neutrino masses and magnitudes of all
of the elements of the leptonic mixing matrix are then predicted in
terms of the single additional parameter $M$.  The successful ansatze
predict a tau neutrino mass in the relevant range for the dark matter
problem and structure formation, muon and electron neutrinos
consistent with the MSW solution to the solar neutrino problem, and
tau-muon neutrino mixing at a level which should soon be observed by
the CHORUS and NOMAD experiments. }

\begin{titlepage}
\begin{flushright}
SHEP 95-28 \\ hep-ph/9509205 \\
\end{flushright}
\vspace{.4in}
\begin{center}
{\large{\bf \sheptitle}}
\bigskip \\ \shepauthor \\ \mbox{} \\ {\it \shepaddress} \\
\vspace{.5in}
{\bf Abstract} \bigskip \end{center} \setcounter{page}{0}
\shepabstract
\end{titlepage}

\section{Introduction}

The Mikheyev-Smirnov-Wolfenstein (MSW) mechanism
\cite{MSWi,MSWii,MSWiii}, which has been proposed to account for the
observed deficits in neutrino flux from the Sun when compared to
standard solar model (SSM) predictions, implies that the electron
neutrino $\nu_{e}$ may mix with another neutrino with a mass
difference and mixing angle given by \cite{KM} $\Delta m^2\approx
3\times 10^{-6} - 1\times 10^{-5}$ eV$^2$, and $\sin^22 \theta \approx
10^{-3}-10^{-2}$, plus a larger angle solution.  The simplest
theoretical interpretation of these data is that the muon neutrino has
a mass $m_{\nu_{\mu}}\approx 2-3\times 10^{-3}$ eV and mixes
dominantly with the electron neutrino which is much lighter. The
status and prospects for confirmation of these results has been
thoroughly discussed elsewhere \cite{HL}.  The theoretical
implications of these results has also been widely discussed in the
literature, most recently within the framework of supersymmetric grand
unified models (SUSY GUTs) \cite{SUSY} based on the idea of a see-saw
mechanism \cite{seesawi,seesawii}.

The see-saw mechanism assumes the neutrino mass terms are of the form
\begin{equation}
\left[ \overline{(\nu_L)}\ \overline{(\nu_R)^c} \right]
\left[\begin{array}{cc}
0 & m_D/2 \\ (m_D)^T/2 & M \\
\end{array}\right] \left[\begin{array}{c}
(\nu_L)^c \\ (\nu_R) \ \\
\end{array}\right] + \mbox{h.c.} =
m_D \overline{(\nu_L)} \ (\nu_R) + M \overline{(\nu_R)^c} \ (\nu_R) +
\mbox{h.c.}  , \label{seesawmatrix}
\end{equation}
where $M >> m_D$ is the Majorana mass of the right handed neutrino.
Such a mass matrix has eigenvalues $m_1\sim {m_D}^2/4M$, $m_2\sim M$.
In general, $m_D$ and $M$ are 3 by 3 matrices in family space and in
order to make predictions for neutrino masses and mixing angles, one
needs to know both $m_D$ and $M$.

A popular back-of-the-envelope suggestion is that
\begin{equation}
m_{\nu_e}:m_{\nu_{\mu}}:m_{\nu_{\tau}}\sim m_u^2:m_c^2:m_t^2
\label{envelope}
\end{equation}
assuming $m_D$ to be given by the corresponding up, charm and top
masses.  Taking equal values of $M\sim 10^{11}$ GeV for the three
families implies $m_{\nu_{\mu}}\approx 2-3\times 10^{-3}$ eV, and
$m_{\nu_{\tau}} \sim 50$ eV, which implies that the tau neutrino may
play a cosmological role as hot dark matter.  However if the neutrino
mixing angle is assumed to be equal to the Cabibbo angle, then one
would obtain $\sin^22 \theta \approx 0.18$ which is not in an MSW
allowed region \cite{HL}.

Clearly the above estimate is too simplistic: it ignores important
theoretical effects due to group theoretical Clebsch coefficients, and
renormalisation group (RG) running which will be important in any
realistic calculation.  However the problem is not just calculational
it is conceptual, since the problem of neutrino masses and mixing
angles is linked to the long standing problem of quark and charged
lepton masses and quark mixing angles.  Furthermore, the above
estimate assumes a relationship between quark and lepton masses which
may or may not be realised in any given theory. Finally the above
estimate makes a simple assumption about the Majorana neutrino mass
matrix, without which it is impossible to make any progress at all.
Faced with these issues theorists have resorted to as many SUSY GUT
models and approaches as there are theoretical papers on the subject,
and no common consensus has yet emerged.  However it is interesting
that several such models
\cite{ELN,so10nonren,nuso10i,nuso10ii}
predict $\nu_{\mu}-\nu_{\tau}$ mixing at levels which are within the
range of sensitivity of the CHORUS \cite{CHORUS}, NOMAD \cite{NOMAD}
and P803 \cite{P803} experiments, which are ideally suited to
$m_{\nu{\tau}}\geq$ a few eV.  The present limit on such mixings of
$\sin^22 \theta\leq 5\times 10^{-3}$ will be pushed down by over an
order of magnitude, providing ample scope for an astonishing
experimental discovery.

Given the existence of neutrino masses, the first deduction we could
make is that the standard model is ruled out!  This would also apply
to those string models which give birth to the SUSY standard model at
the Planck scale. Minimal SU(5) supergravity models would also be
ruled out\footnote{Such models can easily be supplemented by
non-renormalisable operators capable of giving the left-handed
neutrino a Majorana mass, without the introduction of right-handed
neutrinos. In this paper we shall focus only on models which exploit
the see-saw mechanism, however.}  A minimum requirement for a see-saw
mechanism of the type which we are assuming is the existence of
right-handed neutrino fields $\nu_{e_R},\nu_{\mu_R},\nu_{\tau_R}$,
which are singlets under the standard model gauge group.  Having
introduced such fields, it then becomes theoretically possible to
gauge the symmetry $B-L$, where $B$ is baryon number and $L$ is lepton
number, since $B-L$ is now anomaly-free.  This is theoretically very
attractive since gauge symmetries are preferable to accidental global
symmetries, and we find this irresistible. Having gauged $B-L$, it is
natural to assume further unification, such as the SO(10) gauge group
which contains B-L
\cite{so10nonren}.

However, despite its popularity in the literature, there are serious
disadvantages to choosing SO(10) which one should be aware of.  It
turns out that, in order to construct a right-handed neutrino Majorana
matrix, one requires a large Higgs representation such as a
$\underline{126}$ dimensional representation of SO(10). Furthermore,
the scale of right-handed neutrino masses will be set by the grand
unified scale $M_X\approx 10^{16}$ GeV which is much larger than the
desirable scale $M\sim 10^{11}$ GeV encountered earlier.  In order to
rectify this one must rely on either Clebsch factors \cite{so10nonren}
or introduce an intermediate scale of B-L symmetry breaking into the
non-SUSY model\cite{Mohapatra}.  In fact in the non-SUSY version it is
possible to generate suitable Majorana masses without such large Higgs
representations by a two-loop mechanism
\cite{o10neutrinos}, but in SUSY GUTs this mechanism fails
because they are suppressed by the effects of the non-renormalisation
theorem.  In addition SO(10) is subject to the Higgs doublet-triplet
splitting problem, which has no definitive solution yet within the
context of string theory. Of course there are viable solutions to the
doublet-triplet splitting problem available in the literature
\cite{doublet-triplet}
but in the model we shall discuss this problem never arises in the
first place, and we regard this as an attractive feature of our
scheme.

There are alternatives to SO(10) in which right-handed neutrinos are
mandatory and the see-saw mechanism may be exploited.  For example
consider two gauge groups SU(5)$\otimes$U(1) (commonly referred to as
flipped SU(5)) \cite{flippedsu5} and $SU(4)\otimes SU(2)^2$ (which we
shall refer to as 422)
\cite{pati,leo1,leo2,patimass,stevesusy422}.  Both these models are
``string friendly'' in the sense that they do not involve large Higgs
representations, and either do not suffer from a doublet-triplet
splitting problem (in the case of the 422 model) or its resolution is
trivial (in the case of flipped SU(5).)  Neutrino masses have been
studied in both models
\cite{ELN,RP}. In both models the neutrino Majorana matrix
can be generated from higher dimensional operators, leading to a
natural suppression for the scale $M\sim {M_X}^2/M_{Planck}$.  However
in both cases this simple picture is complicated by further singlet
fields (i.e.\ singlets of the full gauge symmetry) which can mix with
the right-handed neutrinos leading to a more complicated neutrino mass
matrix.  In the presence of singlets the physical neutrino masses are
of order ${m_D}^2\mu/{M_X}^2$, where $\mu$ is the mass scale of the
singlets. Clearly if it is assumed that $\mu$ is of order the
electroweak scale then this will result in ultra-small neutrino
masses. However if $\mu \approx M_{Planck}$ then interesting neutrino
masses result.

Recently we have performed a fermion mass operator analysis on the
SUSY 422 model
\cite{patimass}.  According to this analysis the entire spectrum of
quark and charged lepton masses and quark mixing angles can be
reproduced by a suitable set of operators, assuming renormalisable
operators for the third family only and non-renormalisable operators
for the lighter families. In this model the gauged B-L generator is
easily identified as one of the diagonal SU(4) generators, and the
leptons can be regarded as a fourth quark colour\cite{pati}.  Indeed,
many of the predictions valid in SO(10), such as third family Yukawa
unification, and intimate relationships between quark and lepton
masses via Clebsch coefficients are valid in the 422 model.  For
example relations analogous to those in Eq.\ref{envelope} may be valid
in the 422 model but modified by Clebsch and RG
considerations. Analogous relations would certainly not be valid in
flipped SU(5).  Although neutrino masses were not considered in this
analysis, the Dirac part of the neutrino mass matrix is automatically
predicted without further assumption.  In order to obtain the physical
neutrino masses and mixing angles we need to supplement the theory
with the right-handed neutrino Majorana mass matrix.

In the present paper we shall extend the results of the previous
operator analysis to include neutrinos, by including a very simple
ansatz for the right-handed neutrino Majorana mass matrix.  In the
face of complete ignorance about the Majorana matrix, we shall resort
to the simplest possible assumption that can be made, namely that of
universal Majorana neutrino masses.  We shall ignore singlet mixing
completely (which is equivalent to ignoring the singlets completely)
and assume that the right-handed neutrino Majorana matrix is equal to
the three-dimensional unit matrix multiplied by an overall scale
factor $M$. This is a big assumption, but at least it has the virtue
of simplicity, and leads to a neutrino spectrum which is physically
very interesting.  To be specific, we shall find cases where the
neutrino spectrum is of the following form:
\begin{eqnarray}
m_{\nu_e}\ll m_{\nu_{\mu}} & \approx & 2-3\times 10^{-3} \mbox{~eV}
\nonumber \\
\sin^22 \theta_{e\mu} & \approx & 10^{-3}-10^{-2} \nonumber \\
m_{\nu{\tau}} & \geq & 8 \mbox{~eV} \nonumber \\
\sin^22 \theta_{\mu \tau} & \approx & 2-5\times 10^{-3}.
\end{eqnarray}
This spectrum involves an inputted MSW value of the muon neutrino mass
in the above range in order to set the mass parameter $M$. Having
fixed $M$ all the other parameters listed above are true
predictions. Thus the value of the MSW mixing angle, which is clearly
in the correct range, is a genuine prediction of the scheme.  We also
predict the tau neutrino mass to be in the cosmologically interesting
range. Finally our model predicts that CHORUS, NOMAD and P803 will all
see $\nu_{\mu}\rightarrow \nu_{\tau}$ oscillations in the near future.

The layout of the remainder of this paper is as follows: in
section~\ref{sec:model}, the 422 model is introduced, specifically the
superfield gauge irreducible representations and the
superpotential. In section~\ref{sec:ops}, the non-renormalisable
operators employed to provide masses and mixings for the lighter
fermions in ref.\cite{patimass} are reviewed and a summary of the
ansatze that gave successful predictions in the charged fermion sector
is given. In section~\ref{sec:Gen}, the renormalisation group
procedure is extended to include neutrino Yukawa couplings and the
additional heavy mass scale of the right handed tau neutrino
threshold. We discuss the full three family implementation of the
see-saw mechanism, including complex phases in
section~\ref{sec:seesaw}. In section~\ref{sec:pred}, the
diagonalisation of the charged fermion and neutrino Yukawa couplings
is discussed and the resulting relations between fermion masses and
mixing angles are displayed. The predictive ansatze are filtered and
constrained by the MSW effect, the cosmological bound and the E531
data in section~\ref{sec:filter}. Finally, the the prospects for the
model and a summary of the neutrino mass and mixing prediction is
presented in the conclusion, section~\ref{sec:conc}.

\section{The Model\label{sec:model}}
Here we briefly summarise the parts of the model which are relevant
for our analysis.  For a more complete discussion see \cite{leo1}.
The gauge group is,
\begin{equation}
\mbox{SU(4)}\otimes \mbox{SU(2)}_L \otimes \mbox{SU(2)}_R. \label{422}
\end{equation}
The left-handed quarks and leptons are accommodated in the following
representations,
\begin{equation}
{F^i}^{\alpha a}=(4,2,1)=
\left(\begin{array}{cccc}
u^R & u^B & u^G & \nu \\ d^R & d^B & d^G & e^-
\end{array} \right)^i
\end{equation}
\begin{equation}
{\bar{F}}_{x \alpha}^i=(\bar{4},1,\bar{2})=
\left(\begin{array}{cccc}
\bar{d}^R & \bar{d}^B & \bar{d}^G & e^+  \\
\bar{u}^R & \bar{u}^B & \bar{u}^G & \bar{\nu}
\end{array} \right)^i
\end{equation}
where $\alpha=1\ldots 4$ is an SU(4) index, $a,x=1,2$ are
SU(2)$_{L,R}$ indices, and $i=1\ldots 3$ is a family index.  The Higgs
fields are contained in the following representations,
\begin{equation}
h_{a}^x=(1,\bar{2},2)=
\left(\begin{array}{cc}
  {h_2}^+ & {h_1}^0 \\ {h_2}^0 & {h_1}^- \\
\end{array} \right) \label{h}
\end{equation}
(where $h_1$ and $h_2$ are the low energy Higgs superfields associated
with the MSSM.) The two heavy Higgs representations are
\begin{equation}
{H}^{\alpha b}=(4,1,2)=
\left(\begin{array}{cccc}
u_H^R & u_H^B & u_H^G & \nu_H \\ d_H^R & d_H^B & d_H^G & e_H^-
\end{array} \right) \label{H}
\end{equation}
and
\begin{equation}
{\bar{H}}_{\alpha x}=(\bar{4},1,\bar{2})=
\left(\begin{array}{cccc}
\bar{d}_H^R & \bar{d}_H^B & \bar{d}_H^G & e_H^+ \\
\bar{u}_H^R & \bar{u}_H^B & \bar{u}_H^G & \bar{\nu}_H
\end{array} \right). \label{barH}
\end{equation}

The Higgs fields are assumed to develop VEVs,
\begin{equation}
<H>=<\nu_H>\sim M_{X}, \ \ <\bar{H}>=<\bar{\nu}_H>\sim M_{X}
\label{HVEV}
\end{equation}
leading to the symmetry breaking at $M_{X}$
\begin{equation}
\mbox{SU(4)}\otimes \mbox{SU(2)}_L \otimes \mbox{SU(2)}_R
\longrightarrow
\mbox{SU(3)}_C \otimes \mbox{SU(2)}_L \otimes \mbox{U(1)}_Y
\label{422to321}
\end{equation}
in the usual notation.  Under the symmetry breaking in
Eq.\ref{422to321}, the Higgs field $h$ in Eq.\ref{h} splits into two
Higgs doublets $h_1$, $h_2$ whose neutral components subsequently
develop weak scale VEVs,
\begin{equation}
<h_1^0>=v_1, \ \ <h_2^0>=v_2 \label{vevs}
\end{equation}
with $\tan \beta \equiv v_2/v_1$.

In addition to the Higgs fields in Eqs.~\ref{H},\ref{barH} the model
also involves an SU(4) sextet field $D=(6,1,1)$.  The superpotential
of the model is a simplified version\footnote{Without Higgs singlet
fields.} of the one in ref.\cite{leo1}:
\begin{equation}
W=\lambda^{ij}_1F_i\bar{F}_jh
+\lambda_2HHD+\lambda_3\bar{H}\bar{H}D+\mu hh \label{W}
\end{equation}
Note that this is not the most general superpotential invariant under
the gauge symmetry. Additional terms not included in Eq.\ref{W} may be
forbidden by imposing suitable discrete symmetries, the details of
which need not concern us here.  The $D$ field does not develop a VEV
but the terms in Eq.\ref{W} $HHD$ and $\bar{H} \bar{H}D$ combine the
colour triplet parts of $H$, $\bar{H}$ and $D$ into acceptable
GUT-scale mass terms \cite{leo1}. When the $H$ fields attain their
VEVs at $M_X\sim10^{16}$ GeV, the superpotential of Eq.\ref{W} reduces
to the MSSM.\ Note that the last term in Eq.\ref{W} is proportional to
the dimensionful parameter $\mu$. This parameter could be generated by
a gauge singlet field that attains a VEV of order the weak scale, as
in the Next-to-Minimal Supersymmetric Standard Model
(NMSSM)~\cite{NMSSMi,NMSSMii}. In the NMSSM, an extra term must be
added to Eq.\ref{W} but this is not expected to significantly change
any of the results in this paper~\cite{ourbtau}.  Below $M_{X}$ the
part of the superpotential involving matter superfields is just
\begin{equation}
W =\lambda^{ij}_UQ_i\bar{U}_jh_2+\lambda^{ij}_DQ_i\bar{D}_jh_1
+\lambda^{ij}_EL_i\bar{E}_jh_1+ \lambda^{ij}_NL_iN_jh_2 + \ldots
\label{MSSMmatter}
\end{equation}
where $N_i$ are the superfields associated with the right-handed
neutrinos. The Yukawa couplings in Eq.\ref{MSSMmatter} satisfy the
boundary conditions
\begin{equation}
\lambda^{ij}_1 (M_X) \equiv \lambda^{ij}_U(M_X) = \lambda^{ij}_D (M_X)=
\lambda^{ij}_E(M_X) = \lambda^{ij}_N(M_X). \label{boundary}
\end{equation}

\section{Operator Analysis\label{sec:ops}}

In this section, in order to make this paper self-contained, we
briefly review the results of the operator analysis of
ref.\cite{patimass}, extending the resulting tables of Clebsch
coefficients to include the neutrinos.

The boundary conditions listed in Eq.\ref{boundary} lead to
unacceptable mass relations for the light two families. Also, the
large family hierarchy in the Yukawa couplings appears to be unnatural
since one would naively expect the dimensionless couplings all to be
of the same order. This leads us to the conclusion that the
$\lambda^{ij}_1$ in Eq.\ref{W} may not originate from the usual
renormalisable tree level dimensionless coupling.  We allow a
renormalisable Yukawa coupling in the 33 term only and generate the
rest of the effective Yukawa couplings by non-renormalisable operators
that are suppressed by some higher mass scale. This suppression
provides an explanation for the observed fermion mass hierarchy.

We shall restrict ourselves to all possible non-renormalisable
operators which can be constructed from different group theoretical
contractions of the fields:
\begin{equation}
O_{ij}\sim (F_i\bar{F}_j
)h\left(\frac{H\bar{H}}{{M_S}^2}\right)^n+{\mbox H.c.} \label{op}
\end{equation}
where we have used the fields $H,\bar{H}$ in Eqs.\ref{H},\ref{barH}
and $M_S$ is the large scale $M_S>M_{X}$.  The idea is that when $H,
\bar{H}$ develop their VEVs such operators will become effective
Yukawa couplings of the form $h F \bar{F}$ with a small coefficient of
order $M_X^2/M_S^2$.  We shall only consider up to $n=2$ operators
here, since as we shall see even at this level there are a wealth of
possible operators that are encountered.

We shall assume that the Yukawa matrices at $M_{X}$ are all of the
form
\begin{equation}
Y^{U,D,E,N} = \left(
\begin{array}{ccc}
O(\epsilon^2) & O(\epsilon^2) & 0 \\ O(\epsilon^2) & O(\epsilon) &
O(\epsilon) \\ 0 & O(\epsilon) & O(1) \\
\end{array}\right),
\label{matrixform}
\end{equation}
where $\epsilon << 1$ and some of the elements may have approximate or
exact texture zeroes in them. Eq.\ref{matrixform} allows us to
consider the lower 2 by 2 block of the Yukawa matrices first.  In
diagonalising the lower 2 by 2 block separately, we introduce
corrections of order $\epsilon^2$ and so the procedure is consistent
to first order in $\epsilon$.  In a previous paper~\cite{patimass}, we
found several maximally predictive ansatze that were constructed out
of the operators whose Clebsch coefficients are listed in
table~\ref{tab:clebschn1} for the $n=1$ operators. The explicit
operators in component form are listed in the appendices of
ref.\cite{patimass}.
\begin{table}
\begin{center}
\begin{tabular}{|c|c|c|c|c|} \hline
 & $Q \bar{U} h_2$ & $Q \bar{D} h_1$ & $L \bar{E} h_1$ & $L \bar{N}
h_2$
\\ \hline
$O^A$ &1 & 1 & 1 &1 \\ $O^B$ &1 & -1& -1 &1 \\ $O^C$ &1 & 1 & -3 &-3
\\ $O^D$ &1 & -1& 3 &-3\\ $O^G$ &0 & 1 & 2 &0 \\ $O^H$ &2 & 1 & 2 &4\\
$O^K$ &1 & 0 &0 & 3/4\\ $O^M$ &0 & 1 & 1&0 \\ $O^N$ &1&0&0&0\\ $O^O$ &
1&0&0&2\\ \hline
\end{tabular}
\end{center}
\caption{When the Higgs fields develop their VEVs at $M_X$, the
$n=1$ operators utilised lead to the effective Yukawa couplings with
Clebsch coefficients as shown.}
\label{tab:clebschn1}
\end{table}
These $n=1$ operators were used in the lower right hand block of the
Yukawa matrices. We label these successful lower 2 by 2 ansatze $A_i$:
\begin{eqnarray}
A_1 &=& \left[\begin{array}{cc} O^D_{22} - O^C_{22} & 0 \\ O^{B}_{32}
& O_{33} \end{array}\right] \label{A1} \\ A_2 &=&
\left[\begin{array}{cc} 0 & O^A_{23} - O^B_{23} \\ O^D_{32} & O_{33}
\end{array}\right]  \\
A_3 &=& \left[\begin{array}{cc} 0 & O^C_{23} - O^D_{23} \\ O^B_{32} &
O_{33}
\end{array}\right]  \\
A_4 &=& \left[\begin{array}{cc} 0 & O^{C}_{23} \\ O^A_{32} - O^B_{32}
& O_{33}
\end{array}\right]  \\
A_5&=& \left[\begin{array}{cc} 0 & O^{A}_{23} \\ O^C_{32} - O^D_{32} &
O_{33}
\end{array}\right] \\
A_6&=& \left[\begin{array}{cc} O^{K}_{22} & O^{C}_{23} \\ O^M_{32} &
O_{33}
\end{array}\right]\\
A_7&=& \left[\begin{array}{cc} O^{K}_{22} & O^G_{23} \\ O^G_{32} &
O_{33}
\end{array}\right]  \\
A_8&=& \left[\begin{array}{cc} 0 & O^H_{23} \\ O^G_{32} - O^{K}_{32} &
O_{33}
\end{array}\right].
\label{endans}
\end{eqnarray}
Note that replacing $O^K$ with $O^N$ or $O^O$ in $A_{6-8}$ would yield
the same results in the charged fermion sector. The choice of this
operator is however crucial when one considers neutrino masses and so
we denote the choice $A_i^{K},A_i^{N}$ or $A_i^O$, where $i=6,7,8$.

{}From the above ansatze in Eqs.\ref{A1}-\ref{endans}, the ratio of muon
to strange Yukawa couplings at $M_X$ is found to be:
\begin{equation}
\left( \frac{\lambda_\mu}{\lambda_s} \right)_{M_{X}} \equiv l.
\label{2ndpred}
\end{equation}
where $l$ is a ratio of Clebsch coefficients, predicted to be $l=3$,
as in the Georgi-Jarlskog (GJ) \cite{GJ} ansatz or $l=4$ (a new
prediction).  Ansatze $A_{1-6}$ predict $l=3$ in Eq.\ref{2ndpred} and
ansatze $A_{7,8}$ predict $l=4$.

The effective Yukawa couplings generated by the $n=2$ operators have
Clebsch coefficients associated with them, as displayed in
Table~\ref{table:n2}.  Similarly, we now list the $n=2$ operators that
are used to account for $|V_{us}|, m_u, m_d, m_e$ and $|V_{ub}|$.
\begin{table}
\begin{center}
\begin{tabular}{|c|c|c|c|c|} \hline
 & $Q \bar{U} H_2$ & $Q \bar{D} H_1$ & $L \bar{E} H_1$ & $L
\bar{N}H_2$ \\ \hline
$O^{Ad}$ &1 & 3& 9/4 &3/4 \\ $O^{Dd}$ &1 & 3 & 3 &1 \\ $O^{Md}$ &1 &
3& 6 &2\\ $O^1$ &0 & 1 & 1 &0\\ $O^2$ &0 & 1 & 3/4 &0 \\ $O^3$ &0 & 1
& 2 &0 \\ \hline
\end{tabular}
\end{center}
\caption{$n=2$ operators utilised in the upper 2 by 2 ansatze $B_i$.}
\label{table:n2}
\end{table}
The possible ansatze in the down sector that account for correct
$|V_{us}|, m_u, m_d, m_e$ and $|V_{ub}|$ and {\em also} \/generate CP
violation are:
\begin{eqnarray}
B_1 &=& \left[\begin{array}{cc} 0 & O^{n=3}_{12} + O^1_{12}\\
O^{Ad}_{21} & X \end{array}\right]\label{suclighti}\\ B_2 &=&
\left[\begin{array}{cc} 0 & O^{n=3}_{12} + O^2_{12}\\ O^{Ad}_{21} & X
\end{array}\right]\\ B_3 &=& \left[\begin{array}{cc} 0 & O^{n=3}_{12}
+ O^3_{12}\\ O^{Ad}_{21} & X \end{array}\right]\\ B_4 &=&
\left[\begin{array}{cc} 0 & O^{n=3}_{12} + O^1_{12}\\ O^{Dd}_{21} & X
\end{array}\right]\\ B_5 &=& \left[\begin{array}{cc} 0 & O^{n=3}_{12}
+ O^2_{12}\\ O^{Dd}_{21} & X \end{array}\right]\\ B_6 &=&
\left[\begin{array}{cc} 0 & O^{n=3}_{12} + O^3_{12}\\ O^{Dd}_{21} & X
\end{array}\right]\\ B_7 &=& \left[\begin{array}{cc} 0 & O^{n=3}_{12}
+ O^1_{12}\\ O^{Md}_{21} & X \end{array}\right]\\ B_8 &=&
\left[\begin{array}{cc} 0 & O^{n=3}_{12} + O^2_{12}\\ O^{Md}_{21} & X
\end{array}\right]. \label{suclightii}
\end{eqnarray}
$X$ stands for the operator(s) in the 22 position as given earlier.
Each of the successful ansatze, consisting of one of the $A_i$
combined with one of the $B_j$, gives a prediction for the down Yukawa
coupling at $M_X$ in terms of the electron Yukawa coupling:
\begin{equation}
\left( \frac{\lambda_d}{\lambda_e} \right)_{M_{X}} \equiv k,
\label{1stpred}
\end{equation}
where $k=3$ is the GJ prediction of $m_d$. Other viable possibilities
found in our analysis are $k=2,4,\frac{8}{3},\frac{16}{3}$ as shown in
Table~\ref{kvals}.
\begin{table}
\begin{center}
\begin{tabular}{|c|c|c|c|c|c|c|c|c|} \hline
$k$ & $B_1$ & $B_2$ & $B_3$ & $B_4$ & $B_5$ & $B_6$ & $B_7$ & $B_8$ \\
\hline
$A_{1-6}$ & 4 & 16/3 & 2 & 3 & 4 & (3/2) & (3/2) & 2 \\ \hline
$A_{7,8}$ & 16/3 & (64/9) & 8/3 & 4 & 16/3 & 2 & 2 & 8/3 \\ \hline
\end{tabular}
\end{center}
\caption{$k$ values predicted by the ansatze $A_1$ to $A_8$ when
combined with $B_1$ to $B_8$. Note that the bracketed entries predict
$m_d (1$ GeV$)$ to be outside the empirical range.}
\label{kvals}
\end{table}
When the CKM matrix from the ansatze $A_i,B_j$ is calculated, a
prediction of $|V_{ub}|$ in terms of an unknown complex phase $\phi$
is found:
\begin{equation}
|V_{ub}(m_t)|\sim \frac{|V_{us}(m_t)| |V_{cb}(m_t)|}{\sqrt{1 + \left(
\frac{3 \lambda_c(M_X)}{\lambda_s(M_X)} \right)^2 - 2 \cos\phi(M_X)
\frac{3 \lambda_c(M_X)}{\lambda_s(M_X)} } }.
\label{Vubpredict}
\end{equation}
More details of this analysis can be found in ref.\cite{patimass}.

\section{Generalisation to Include Neutrino Masses \label{sec:Gen}}
This analysis of charged fermion masses does not include neutrinos,
even though the model predicts Dirac neutrino masses to be of order
the mass of the up quarks. If this was the case, direct upper neutrino
mass bounds would be violated and so some mechanism must be employed
to suppress them significantly. In this paper, we assume that the
neutrino masses are suppressed via the popular see saw mechanism, as
in Eq.\ref{seesawmatrix}. Since the Dirac masses $m_D$ of $\nu_e$ and
$\nu_\mu$ are several orders of magnitude smaller than that of
$\nu_{\tau}$, it is a good approximation to consider the third family
alone and drop smaller Yukawa couplings, as we did for the charged
fermions.

As shown in ref.~\cite{ourquadun}, the tau neutrino Yukawa coupling
enters the system of renormalisation group equations (RGEs) of the
third family Yukawa couplings. However, the effects of this coupling
on the predictions of triple Yukawa unification
\cite{so10nonren} are negligible compared to the effects of the
experimental uncertainties on $m_b$ and $\alpha_S (M_Z)$. Thus the
numerical predictions of $m_d,m_s,m_b,\tan \beta,m_t$ and $|V_{ub}|$
made in ref.\cite{patimass} still approximately hold once the
neutrinos have been taken into account. At the scale $M$, the right
handed tau neutrino decouples from the effective field theory and so
if one assumes that $M \sim 10^{16}$ GeV, the predictions in
ref.\cite{patimass} hold completely. Such a high value of the Majorana
mass scale would mean very light neutrinos of masses $< O (10^{-3})$
eV, which could not account for any of the solar neutrino, dark matter
or structure formation problems in cosmology.  On the contrary,
Majorana masses of order $10^{10}-10^{12}$ GeV can address these
problems \cite{ourquadun} and it is in this region of the parameter
space that we are going to analyse the ansatze $A_i,B_j$ to obtain
neutrino masses and leptonic mixing angles.

The superpotential in Eq.\ref{W} was used to calculate the running of
the Yukawa couplings in the MSSM and the $\overline{MS}$
renormalisation scheme, using an analysis of general superpotentials
performed by Martin and Vaughn \cite{MandV}:
\begin{eqnarray}
\frac{\partial g_i}{\partial t} &=& \frac{b_i g_i^3}{16 \pi^2}
\nonumber \\
\frac{\partial Y^U}{\partial t} &=& \frac{Y^U}{16 \pi^{2}} \left[
\mbox{Tr}
\left( 3 Y^U {Y^U}^{\dagger} + {Y^N}^\dagger Y^N \right) + 3
{Y^U}^{\dagger} Y^U + {Y^D}^{\dagger} {Y^D} - \left( \frac{13}{15}
g_{1}^{2} + 3g_{2}^{2} + \frac{16}{3} g_{3}^{2} \right )\right]
\nonumber \\
\frac{\partial {Y^D}}{\partial t} &=& \frac{{Y^D}}{16 \pi^{2}} \left[
\mbox{Tr}
\left( 3{Y^D} {Y^D}^{\dagger} + {Y^E} {Y^E}^{\dagger} \right) +
{Y^U}^{\dagger} Y^U + 3 {Y^D}^{\dagger} {Y^D} - \left( \frac{7}{15}
g_{1}^{2} + 3 g_{2}^{2} + \frac{16}{3} g_{3}^{2} \right) \right]
\nonumber \\
\frac{\partial {Y^E}}{\partial t} &=& \frac{{Y^E}}{16 \pi^{2}} \left[
\mbox{Tr}
\left( 3 {Y^D} {Y^D}^{\dagger} + {Y^E} {Y^E}^{\dagger} \right) + 3
{Y^E}^{\dagger} {Y^E} + {Y^N}^\dagger {Y^N} - \left( \frac{9}{5}
g_{1}^{2} + 3 g_{2}^{2} \right)
\right] \nonumber  \\
\frac{\partial Y^N}{\partial t} &=& \frac{Y^N}{16 \pi^2} \left[
\mbox{Tr} \left( 3 {Y^U}^\dagger Y^U + {Y^N}^\dagger Y^N \right)
 + 3 {Y^N}^\dagger Y^N + {Y^E}^\dagger Y^E - \left(
\frac{3}{5}g_1^2 + 3 g_2^2 \right) \right]
, \label{RG1nuR}
\end{eqnarray}
where $b_i=(33/5,1,-3)$, $t = \ln \mu$ and $\mu$ is the
$\overline{MS}$ renormalisation scale.  Once the small couplings have
been dropped, Eqs.\ref{RG1nuR} reduce to the RGEs derived in
\cite{nubtaumass}:
\begin{eqnarray}
16 \pi^2 \frac{\partial g_i}{\partial t} &=& b_i g_i^3 \nonumber \\ 16
\pi^{2} \frac{\partial \lambda_{t}}{\partial t} &=& \lambda_{t}
\left[ 6\lambda_{t}^{2} +  \lambda_{b}^{2}  + \theta_R
\lambda_{\nu_\tau}^2 -
\left( \frac{13}{15}
g_{1}^{2} + 3g_{2}^{2} + \frac{16}{3}g_{3}^{2} \right)
\right] \nonumber \\
16 \pi^{2} \frac{\partial \lambda_{b}}{\partial t} &=&
\lambda_{b} \left[ 6
\lambda_{b}^{2} + \lambda_{\tau}^{2} + \lambda_{t}^{2} -
\left(
\frac{7}{15} g_{1}^{2} + 3g_{2}^{2} +
\frac{16}{3} g_{3}^{2} \right) \right] \nonumber \\
16 \pi^{2} \frac{\partial \lambda_{\tau}}{\partial t} &=&
\lambda_{\tau}
\left[ 4 \lambda_{\tau}^{2} + 3 \lambda_{b}^{2} + \theta_R
\lambda_{\nu_\tau}^2
- \left( \frac{9}{5} g_{1}^{2} + 3g_{2}^{2} \right) \right] \nonumber
\\ 16 \pi^{2} \frac{\partial \lambda_{\nu_\tau}}{\partial t} &=&
\lambda_{\nu_\tau}
\left[ 4 \theta_R \lambda_{\nu_\tau}^{2} + 3 \lambda_{t}^{2} +
\lambda_{\tau}^2
- \left( \frac{3}{5} g_{1}^{2} + 3g_{2}^{2} \right) \right],
\label{3rdfamnuReqs}
\end{eqnarray}
where $\theta_R \equiv \theta(t - \ln M)$ takes into account the large
mass suppression of the right-handed neutrino loops at scales
$\mu<M$. Thus we integrate out loops involving right-handed neutrinos
at $M$, but retain the Dirac Yukawa coupling $\lambda_{\nu_{\tau}}$
which describes the coupling of left to right-handed neutrinos.  The
running procedure to determine the low energy masses is then to run
down the (Dirac) neutrino Yukawa coupling $\lambda_{\nu_{\tau}}$ from
$M_X$ to low-energies using the above RGEs.

In dealing with the first and second families we have to confront the
problem that the Yukawa matrices are not diagonal. As discussed widely
elsewhere \cite{so10nonren,RRR}, it is most convenient to diagonalise
the Yukawa matrices at $M_X$ before running them down to $m_t$. It is
then possible to obtain RGEs for both the {\em diagonal}
\/Yukawa couplings $\lambda_{u,c,t}$, $\lambda_{d,s,b}$,
$\lambda_{e,\mu,\tau}$ and the Cabibbo-Kobayashi-Maskawa (CKM) matrix
elements $|V_{ij}|$\footnote{The empirical values of $|V_{ij}|$ were
taken to be at $m_t$ instead of $M_Z$, introducing an error whose
magnitude is always less than 1 percent for our analysis.}
(ref.\cite{ourbtau,RGstudy,btaunewi,btaunewiii,BandB}). At one-loop
these RGEs can be numerically integrated so that the low energy
physical couplings have a simple scaling behaviour
\begin{eqnarray}
\left( \frac{\lambda_{u,c}}{\lambda_{t}} \right)_{m_t} & = &
\left( \frac{\lambda_{u,c}}
{\lambda_{t}} \right)_{M_{X}} e^{3I_{t} + I_{b}} \label{start}
\nonumber \\
\left( \frac{\lambda_{d,s}}{\lambda_{b}} \right)_{m_t} & = &
\left( \frac{\lambda_{d,s}}
{\lambda_{b}} \right)_{M_{X}} e^{3I_{b} + I_{t}} \nonumber \\
\left( \frac{\lambda_{e, \mu}}{\lambda_{\tau}} \right)_{m_t} & =
& \left(
\frac{\lambda_{e, \mu}}
{\lambda_{\tau}} \right)_{M_{X}} e^{3I_{\tau} + I_{\nu_\tau}}
\nonumber \\
\left( \frac{\lambda_{{\nu_e}, {\nu_\mu}}}{\lambda_{\nu_\tau}}
\right)_{m_t} & =
& \left(
\frac{\lambda_{{\nu_e}, {\nu_\mu}}}
{\lambda_{{\nu_\tau}}} \right)_{M_{X}} e^{I_{\tau} +3I_{\nu_\tau}}
\nonumber \\
\frac{ \mid V_{cb} \mid _{M_{X}}}{\mid V_{cb} \mid
_{m_t} } & = & e^{I_{b}+I_{t}} \nonumber \\
\frac{ \mid V_{\mu \tau} \mid _{M_{X}}}{\mid V_{\mu \tau} \mid
_{m_t} } & = & e^{I_{\tau}+I_{\nu_\tau}},
\label{gutsusy}
\end{eqnarray}
with identical scaling behaviour to $V_{cb}$ of $V_{ub}$, $V_{ts}$,
$V_{td}$. The leptonic mixing elements $V_{e \tau},V_{\tau e},V_{\tau
\mu}$ scale identically to $V_{\mu \tau}$.The $I$ integrals are
defined as
\begin{equation}
I_{i} \equiv \int _{\Lambda_i}^{\ln {M_{X}}} \left( \frac{\lambda_{i}
\left( t
\right) }{4 \pi} \right)^2 dt, \label{iint}
\end{equation}
where $\Lambda_{\nu_\tau}=M$, $\Lambda_{\tau,b,t}=m_t$ and $t=\ln
\mu$, $\mu$ being the $\overline{MS}$ scale.  To a consistent level of
approximation $V_{e \mu}$, $V_{e e}$, $V_{\mu
\mu}$, $V_{\mu e}$, $V_{\tau \tau}$, $V_{us}$, $V_{ud}$, $V_{cs}$,
$V_{cd}$, $V_{tb}$, $\lambda_{\nu_e} /
\lambda_{\nu_\mu}$,$\lambda_{u}$/$\lambda_{c}$,
$\lambda_{d}$/$\lambda_{s}$ and $\lambda_{e}$/$\lambda_{\mu}$ are RG
invariant.  The CP violating quantity J scales as $V_{cb}^{2}$.

\section{Implementing the See-Saw Mechanism\label{sec:seesaw}}
For simplicity, and to make the scheme as predictive as possible, we
assume the Majorana matrix $M$ to be proportional to the unit matrix
in family space. $M$ may be generated by non-renormalisable terms of
the form
\begin{equation}
\lambda \bar{F}_{\alpha x}^j \bar{F}_{\beta y}^j \frac{H^{\alpha x}
H^{\beta y}}{M_{S_1}} \left( \frac{H^{\gamma z} \bar{H}_{\gamma
z}}{M_{S_2}} \right)^m,
\end{equation}
where $M_{S_i}>M_X, i=1,2$ are higher scales\footnote{For example the
compactification scale, the string scale or the Planck scale.}. We do
not explicitly specify the exact operator responsible for $M$, but
merely use a numerical value of $10^{10}-10^{12}$ GeV as a starting
point to solve some of the problems associated with neutrino masses.

The diagonalisation procedure to transform to the mass basis of the
neutrinos seems more complicated than a similar calculation on the
charged fermions because of the additional feature of the Majorana
mass matrix. We now show how the neutrino masses may be diagonalised
in our scheme. We take the mass matrix from Eq.\ref{seesawmatrix},
generalise to the three family case and diagonalise the Dirac masses
by unitary transformations upon the neutrino fields at the unification
scale $M_X$:
\begin{equation}
\left[ \overline{(\nu_L)}_i\ \overline{(\nu_R)^c}_i \right](V^\dag V)^T
\left[ \begin{array}{cc} 0 & m_D/2 \\ {m_D}^T/2 & M \\
\end{array}\right] V^\dag V
\left[\begin{array}{c}
(\nu_L)^c_i \\ (\nu_R)_i \ \\
\end{array}\right] + \mbox{H.c.},
\end{equation}
where each 6 by 6 matrix has been split up into four 3 by 3
submatrices and
\begin{equation}
V\equiv\left[ \begin{array}{cc} V_{N_L}^* & 0 \\ 0 & V_{N_R} \\
\end{array}
\right].
\end{equation}
We redefine the neutrino superfields as $(\nu_L)_i
\rightarrow  V_{N_L}^{ij} (\nu_{L})_j$ and $(\nu_R)_i
\rightarrow  V_{N_R}^{ij} (\nu_{R})_j$ in a basis where $m_D
\rightarrow V_{N_L} m_D V_{N_R}^\protect\dag$ is diagonal. We denote
the diagonalised entries of $m_D$ as $m_i, (i=1\ldots3)$. Note that
the Majorana matrix $M
\rightarrow V_{N_R}^* M V_{N_R}^\dag = M \mbox{diag} (e^{i \phi_e},e^{i
\phi_\mu}, e^{i \phi_\tau})$ (with the last equality following
from the assumption that $M$ is proportional to the unit matrix) now
involves complex phases in general.  The $(\nu_{R})_i$ superfields are
redefined (to a primed basis) to absorb the complex phases present in
the Majorana mass terms
\begin{equation}
M \sum_{i=e,\mu,\tau} e^{i\phi_i} \overline{(\nu_R)^c}_i (\nu_R)_i =M
\sum_{i=e,\mu,\tau} \overline{(\nu'_R)^c}_i (\nu'_R)_i
\end{equation}
where
\begin{equation}
{\nu_{e,\mu,\tau}}_R' = e^{i
\phi_{e,\mu,\tau}/2} {\nu_{e,\mu,\tau}}_R,  \end{equation}
and has the effect of making the neutrino Dirac entries carry phases,
which may then be absorbed into a redefinition of the left handed
neutrino phases
\begin{equation}
(\nu_L)_i' = e^{-i \phi_i/2} (\nu_L)_i\label{LHphase}
\end{equation}
leaving all of the neutrino masses real.  The transformation in
Eq.\ref{LHphase} will produce phases in the leptonic CKM mixing matrix
$V_{LEP}$, which will be defined in the next section.  However it will
turn out that these phases contribute to the overall phase of
individual matrix elements ${V_{LEP}}_{ij}$ and because our
predictions are only in terms of the moduli of these angles, they do
not enter our calculations.  In fact it is easy to see that, because
of our assumption that the Majorana matrix is proportional to the unit
matrix, ${|V_{LEP}|}_{ij}$ may be calculated just from the Dirac parts
of the leptonic matrices ignoring the see-saw mechanism completely.

Thus the strategy is to diagonalise the Dirac Yukawa matrices at
$M_X$, and obtain the diagonal Yukawa couplings and quark and lepton
CKM matrix elements at the scale $M_X$.  These parameters are then run
to low energies using the procedure outlined in section~\ref{sec:Gen},
i.e.\ taking into account the effects of the right handed tau
neutrino. The Majorana masses do not run to one loop.  At low energy
$(\mu=M_W)$, in order to obtain the light neutrino basis, we do
another transformation upon the neutrino mass matrix that performs the
see saw action.  To first order in the small see-saw angle matrix
$\Theta_i=\mbox{diag}(\theta_1,\theta_2,\theta_3)$,
\begin{equation}
\left[ \overline{(\nu_L')}_i\ \overline{(\nu_R')^c}_i \right]U U^T
\left[ \begin{array}{cc} 0 & m_i/2 \\ m_i/2 & M \\ \end{array} \right]
 U U^T
\left[\begin{array}{c}
(\nu_L')^c_i \\ (\nu_R')_i \ \\
\end{array}\right]  + \mbox{H.c.}
\label{seedown}
\end{equation}
where
\begin{equation}
U \equiv \left[ \begin{array}{cc} 1 &-\Theta_i \\ \Theta_i & 1 \\
\end{array}
\right].
\end{equation}
The mass matrix in Eq.\ref{seedown} has three light eigenvalues (the
physical neutrinos) of magnitude $m_{\nu_i} \sim -m_i^2/4M$ and three
heavy eigenvalues $\sim M$. The neutrino mixing angles are $\Theta_i
\sim \mbox{diag}(-m_1/2M,-m_2/2M,-m_3/2M)$ so that the whole 6 by 6
neutrino mass matrix is now diagonal.

\section{Predictions of Leptonic CKM Matrix Elements \label{sec:pred}}

The successful ansatze consist of any of the lower 2 by 2 blocks $A_i$
combined with any of the upper 2 by 2 blocks $B_i$, subject to the
restrictions shown in Table~\ref{kvals}.  For example let us consider
$A_1$ in the lower 2 by 2 block combined with any of the $B_i$ in the
upper 2 by 2 block, focusing particularly on $A_1$ combined with
$B_1$. Just above $M_X$, before the $H,\bar{H}$ fields develop VEVs,
we have the operators
\begin{equation}
\left[\begin{array}{ccc}
0 & O^1_{12} + {O}_{12}^{n=3} & 0 \\ O^{Ad}_{21} & O^D_{22} -
{O}^C_{22} & 0 \\ 0 & O^B_{32} & O_{33} \\ \end{array}\right],
\end{equation}
which implies that at $M_X$ the Yukawa matrices are of the form
\begin{equation}
Y^{U,D,E,N} =
\left[
\begin{array}{ccc}
0 & H_{12}x_{12}^{U,D,E,N} e^{i \phi_{12}}+H_{12}{'}
x{'}_{12}^{U,D,E,N} e^{i \phi_{12}'}& 0 \\ H_{21}x_{21}^{U,D,E,N} e^{i
\phi_{21}} & H_{22}x_{22}^{U,D,E,N} e^{i
\phi_{22}} -
H_{22}' x_{22}'^{U,D,E,N} e^{i \phi_{22}'} & 0 \\ 0 & H_{32}
x_{32}^{U,D,E,N} e^{i \phi_{32}} & H_{33}e^{i \phi_{33}} \\
\end{array}\right],
\label{firstyuk}
\end{equation}
where we have factored out the phases of the operators and $H_{ik}$
are the magnitudes of the coupling constant associated with
$O_{ik}$. Note the real Clebsch coefficients\footnote{See appendix.}
$x_{ik}^{U,D,E,N}$ give the splittings between $Y^{U,D,E,N}$.  We now
make the transformation in the 22 element of Eq.\ref{firstyuk}
\begin{equation}
x_{22}^{U,D,E,N} H_{22} e^{i \phi_{22}} - x{'}_{22}^{U,D,E,N} H_{22}'
e^{i \phi'_{22}}
\equiv H_{22}^{U,D,E,N} e^{i \phi_{22}^{U,D,E,N}},
 \label{transform22}
\end{equation}
where $H_{22}^{U,D,E,N}, \phi_{22}^{U,D,E,N}$ are real positive
parameters. It follows from the Clebsch structure in Eq.\ref{clebsch2}
that $H_{22}^E = 3 H_{22}^D$, $H_{22}^N=3 H_{22}^U$ and $\phi_{22}^E =
\phi_{22}^D$, $\phi_{22}^N = \phi_{22}^U$. In
general we shall write $H_{22}^E = l H_{22}^D$, $H_{22}^N = l_N
H_{22}^U$, where $l=3, l_N=3$ in this case.

We rotate the phases of the $F, \bar{F}$ fields as in the appendix in
order to decrease the number of phases in the Yukawa matrices and
hence derive real mass eigenstates eventually.  Below $M_X$, the
multiplets $F, \bar{F}$ are no longer connected by the gauge symmetry
in the effective field theory, since it is The Standard Model.  We now
define our notation as regards the effective field theory below $M_X$
as follows. The effective quark Yukawa terms are written (suppressing
all indices)
\begin{equation}
(U_R)^c Y^U Q_L h_2 + (D_R)^c Y^D Q_L h_1 + \mbox{H.c.}
\label{quarkL}
\end{equation}
We transform to the quark mass basis by introducing four 3 by 3
unitary matrices $V_{U_{L,R}}, V_{D_{L,R}}$ then the Yukawa terms
become
\begin{equation}
(U_R)^c V_{U_R}^\dagger V_{U_R} Y^U V_{U_L}^\dagger V_{U_L} Q_L h_2 +
(D_R)^c V_{D_R}^\dagger V_{D_R} Y^D V_{D_L}^\dagger V_{D_L} Q_L h_1 +
\mbox{H.c.}
\label{massbasis}
\end{equation}
where $Y^U_{\mbox{diag}} = V_{U_R} Y^U V_{U_L}^\dagger$ and
$Y^D_{\mbox{diag}} = V_{D_R} Y^D V_{D_L}^\dagger$ are the diagonalised
Yukawa matrices.  With the definitions in
Eqs.\ref{quarkL},\ref{massbasis}, the CKM matrix is of the form
\begin{equation}
V_{CKM} \equiv V_{U_L} V_{D_L}^\dagger. \label{CKM}
\end{equation}
The effective lepton Yukawa terms below $M_X$ are defined to be
(suppressing all indices)
\begin{equation}
(\nu_R)^c Y^N L_L h_2 + (e_R)^c Y^E L_L h_1 + \mbox{H.c.}
\end{equation}
We transform to the lepton dirac mass basis by introducing the four 3
by 3 unitary matrices $V_{N_{L,R}}, V_{E_{L,R}}$ and the Yukawa terms
become
\begin{equation}
(\nu_R)^c V^\dag_{N_R} V_{N_R} Y^N V^\dag_{N_L} V_{N_L} L_L h_2 +
(e_R)^c V_{E_R}^\dag V_{E_R} Y^E V_{E_L}^\dag V_{E_L} L_L h_1 +
\mbox{H.c.}
\end{equation}
where $Y^N_{\protect\mbox{diag}}=V_{N_R} Y^N V_{N_L}^\dag$ and
$Y^E_{\protect\mbox{diag}} = V_{E_R} Y^E V_{E_L}^\dag$ are the
diagonalised lepton Yukawa matrices.  The argument in
section~\ref{sec:seesaw} shows that the assumption of a Majorana
matrix proportional to 1 allows us to be only concerned with the
diagonalisation of the Dirac mass matrices to calculate the magnitudes
of the leptonic mixing matrix elements
\begin{equation}
|V_{LEP}|_{ij} \equiv |(V_{N_L})_{ik}
(V_{E_L}^\dag)_{kj}|. \label{LEP}
\end{equation}

In all of the cases considered, $x_{12}^U=0$ and $x_{12}'^{D,E}=0$ so
that the Yukawa matrices which result from
Eqs.\ref{firstyuk},\ref{transform22},\ref{phaserots} are
\begin{eqnarray}
Y^D &=& \left[\begin{array}{ccc} 0 & H_{12}x_{12}^D & 0 \\ H_{21}
x_{21}^D & H_{22}^D& 0 \\ 0 & H_{32}x_{32}^D & H_{33} \\
\end{array}\right] \nonumber \\
Y^E &=& \left[\begin{array}{ccc} 0 & H_{12}x_{12}^E & 0 \\ H_{21}
x_{21}^E & l H_{22}^D & 0 \\ 0 & H_{32}x_{32}^E & H_{33} \\
\end{array}\right] \nonumber \\
Y^U &=& \left[\begin{array}{ccc} 0 & H_{12}'{x'}^U_{12} e^{i
(\phi_{12}{'} - \phi_{12})} & 0 \\ x_{21}^U H_{21} & H_{22}^U e^{i
(\phi_{22}^U- \phi_{22}^D)} & 0 \\ 0 & H_{32} x_{32}^U & H_{33} \\
\end{array}\right] \nonumber \\
Y^N &=& \left[\begin{array}{ccc} 0 & 0 & 0 \\ x_{21}^N H_{21} & l_N
H_{22}^U e^{i (\phi_{22}^U- \phi_{22}^D)} & 0 \\ 0 & H_{32} x_{32}^N &
H_{33} \\
\end{array}\right]. \label{expivi}
\end{eqnarray}
Note that in Eq.\ref{expivi}, we have assumed that the $O^{n=3}_{12}$
operator doesn't give a Yukawa term to the neutrino for simplicity.
None of the predictions change if this limit is revoked except that
the electron neutrino is approximately massless rather than exactly
massless.  In order to diagonalise the quark Yukawa matrices, we first
make $Y^{U,N}$ real, by multiplying by phase matrices, as in the
appendix.

The diagonal Yukawa couplings of the strange quark and muon obtained
from Eq.\ref{diags} are $(\lambda_s)_{M_X}=H_{22}^D$ and
$(\lambda_\mu)_{M_X} =l H_{22}^D$ since the 22 eigenvalues are just
the 22 elements in this case.  Similarly, we use
$(\lambda_c)_{M_X}=H_{22}^U$ and $(\lambda_{\nu_\mu})_{M_X}=l_N
H_{22}^U$.  This gives us the prediction of the strange quark and
neutrino Yukawa couplings
\begin{equation}
\begin{array}{cc}
\lambda_s(M_X) = \lambda_\mu(M_X) / l, & \lambda_{\nu_\mu} = l_N
\lambda_{c}. \\ \end{array} \label{2famprd}
\end{equation}
$A_6$ makes no such prediction and so we discard it on the grounds
that we are searching for the most predictive (and therefore testable)
cases. $A_7^N, A_8^N$ are special cases because they predict a
massless muon neutrino. We consider these cases later.

The first family diagonal Yukawa couplings for the down quark and
electron are related by
\begin{equation}
\left( \frac{\lambda_d}{\lambda_e} \right)_{M_X}= l \frac{x_{21}^D
x_{12}^D}{x_{21}^E x_{12}^E}.
\label{givesk}
\end{equation}
We identify the right hand side of Eq.\ref{givesk} with $k$ in
Eq.\ref{1stpred}.  We now substitute the diagonalising matrices from
Eqs.\ref{diags} and~\ref{phasegone} into the CKM matrix in
Eq.\ref{CKM} to obtain
\begin{equation}
|V_{CKM}|_{ij} = \left[\begin{array}{ccc} c_2 c_1 e^{i \phi} + s_2 s_1
c_3 & -s_2 c_1 e^{i \phi} + s_1 c_2 c_3 & s_1 s_3 \\ -s_1 c_2 e^{i
\phi} + c_1 s_2 c_3 & s_1 s_2 e^{i \phi} + c_2 c_3 c_1 & s_3 c_1 \\
-s_2 s_3 & -c_2 s_3 & c_3 \end{array}\right]. \label{genmix}
\end{equation} Substituting Eqs.~\ref{diags} and~\ref{phasegone} into
the leptonic mixing matrix in Eq.~\ref{LEP} yields the matrix
$V_{LEP}$ form as $V_{CKM}$, except with all angles $\theta_i$
replaced by their leptonic analogues $\theta_i^l$, defined as in the
appendix.

We may start making predictions in the leptonic sector at $M_X$ for
the elements of $|V_{LEP}|$, in terms of those of $|V_{CKM}|$.
Proceeding in the manner laid out in the appendix, we obtain
\begin{eqnarray}
|V_{\mu \tau}(M_X)|&=&m |V_{cb}(M_X)| \label{mutaupred} \\ |V_{e
\tau}(M_X)| &=& n |V_{ub}(M_X)| \label{etauub} \\ |V_{\tau e}(M_X)|
&=& o |V_{ub}(M_X)| \frac{\lambda_c}{\lambda_\mu}
\label{taue} \\
|V_{e \mu}(M_X)| &=& p \frac{|V_{ub}(M_X)|}{|V_{cb}(M_X)|}
\sqrt{\left( \frac{q \lambda_c(M_X)}{\lambda_\mu(M_X)} \right)^2 - 2
q\frac{\lambda_c(M_X)}{\lambda_{\mu(M_X)}} \cos \phi + 1},
\label{emupred}
\end{eqnarray}
where $m,n,o,p,q$ are all rational numerical Clebsch coefficients, as
shown in the appendix. The uncertainty due to the arbitrary phase
$\phi$ gives a range of values for $|V_{e \mu}|$, even if all of the
other parameters are fixed.  However, it was seen in
ref.\cite{patimass} that the values of $|V_{ub}|$ predicted by
Eq.\ref{Vubpredict} are only within the experimentally measured range
for $\phi > \pi/2$. When making predictions, we therefore examine the
endpoints of the valid range, that is $\cos
\phi=0, -1$.

The diagonalisation procedure in the two special cases $A_7^N,A_8^N$
is somewhat simpler since the neutrino Yukawa matrix consists of only
two non-zero entries. $A_8^N$ is discarded because it does not make a
prediction for $|V_{\mu \tau}|$. When $A_7^N$ combined with any of the
$B_i$ ansatze is diagonalised, the leptonic mixing matrix is
\begin{equation}
V_{LEP} = \left[\begin{array}{ccc} c_2^l & -s_2^l & 0 \\
\bar{c}_4^l s_2^l & c_2^l \bar{c}_4^l & -\bar{s}_4^l \\
0 & c_2^l \bar{s}_4^l & \bar{c}_4^l \\ \end{array}\right],
\end{equation}
leading to the predictions
\begin{equation}
\begin{array}{c} \begin{array}{ccc}
|V_{\mu \tau}(M_X)|=2 |V_{cb}(M_X)|,& |V_{\tau e}(M_X)|=0,& |V_{e
\tau}(M_X)|=0,\\ \end{array} \\ \begin{array}{cc}|V_{e \mu}(M_X)|=
-\frac{x_{21}^E}{x_{21}^U} \frac{\lambda_c(M_X)}{\lambda_\mu (M_X)}
\frac{|V_{ub}(M_X)|}{|V_{ub}(M_X)|}, & \lambda_{\nu_\mu}(M_X) =
\frac{|V_{ub}(M_X)|}{|V_{cb}|(M_X)} \lambda_c(M_X) \\ \end{array} \\
\end{array}.
\label{oddpred}
\end{equation}

Table~\ref{ClLEPmix} displays all of the Clebsch coefficients
associated with the predictions in
Eqs.\ref{2famprd},\ref{mutaupred}-\ref{emupred}.
\begin{table}
\begin{center}
\def\arraystretch{1.5}
\begin{tabular}{|c|c|c|c|} \hline
$(l_N,m,n,o,p,q)$ & $B_{1,2,3}$ & $B_{4,5,6}$ & $B_{7,8}$ \\ \hline
$A_{1,3,4}$ & (3,1,$\frac{1}{4}$,$\frac{9}{4}$,$\frac{1}{4}$,9) &
(3,1,$\frac{1}{3}$,3,$\frac{1}{3}$,9) &
(3,1,$\frac{2}{3}$,6,$\frac{2}{3}$,9) \\
$A_{2,5}$ & (3,3,$\frac{3}{4}$,$\frac{27}{4}$,$\frac{1}{4}$,9) &
(3,3,1,9,$\frac{1}{3}$,9) & (3,3,2,18,$\frac{2}{3}$,9) \\
$A_7^K$ & ($\frac{3}{4}$,2,2,$\frac{9}{2}$,1,$\frac{9}{4}$) &
($\frac{3}{4}$,2,$\frac{8}{3}$,6,$\frac{4}{3}$,$\frac{9}{4}$) &
($\frac{3}{4}$,2,$\frac{16}{3}$,12,$\frac{8}{3}$,$\frac{9}{4}$)\\
$A_7^O$ & (2,2,$\frac{3}{4}$,$\frac{9}{2}$,$\frac{3}{8}$,6) &
(2,2,1,6,$\frac{1}{2}$,6) & (2,2,2,12,1,6) \\
$A_8^K$& ($\frac{3}{2}$,$\frac{3}{4}$,$\frac{3}{8}$,$\frac{27}{16}$,
$\frac{1}{2}$,$\frac{9}{2}$)&
($\frac{3}{2}$,$\frac{3}{4}$,$\frac{1}{2}$,$\frac{9}{4}$,$\frac{2}{3}$,$\frac{9}{2}$)
&
($\frac{3}{2}$,$\frac{3}{4}$,1,$\frac{9}{2}$,$\frac{4}{3}$,$\frac{9}{2}$)\\
$A_8^O$& (4,2,$\frac{3}{8}$,$\frac{9}{2}$,$\frac{3}{16}$,12) &
(4,2,$\frac{1}{2}$,6,$\frac{1}{4}$,12) & (4,2,1,12,$\frac{1}{2}$,12)\\
\hline
\end{tabular}
\end{center}
\caption{Table of Clebsch coefficients for the lepton mass and mixing
angle predictions defined in
Eqs.\protect\ref{2famprd},\protect\ref{mutaupred}-\protect\ref{emupred}.}
\label{ClLEPmix}
\end{table}
Having constrained the leptonic mixing matrix elements at $M_X$ for
any particular ansatz $A_i,B_j$ with the predictions in
Eqs.\ref{emupred}-\ref{mutaupred}, the RGEs in Eq.\ref{gutsusy} are
employed to yield low-energy predictions of the quantities. The mixing
angle predictions are displayed in Table~\ref{LEPmixi}.  Note that the
mixing angles displayed in Table~\ref{LEPmixi} are not significantly
dependent upon $M$ when compared to the uncertainties in the
predictions. Part of the uncertainties of the predictions within one
particular ansatz $A_i,B_j$ are correlated by $\alpha_S(M_Z),m_b$.
For reasons of brevity, we refrain from displaying these correlations
for all possible ansatze.  The table illustrates the powerful
predictability (and therefore testability) of our scheme.
\begin{table}
\begin{center} \begin{tabular}{c}
  \begin{tabular}{|c|c|c|c|c|c|c|} \hline $B_{1,2,3}$ &
 $A_{1,3,4}$&$A_{2,5}$&$A_7^K$&$A_7^O$&$A_8^K$&$A_8^O$ \\ \hline
 $|V_{\mu
 \tau}|/10^{-1}$&0.32-0.55&0.95-1.64&0.63-1.09&0.63-1.09&0.24-0.41&0.63-1.09\\
 $|V_{e\mu}|/10^{-2}$&1.6-4.6&1.6-4.6&6.0-12.1&2.3-5.8&3.0-7.1&1.2-3.9\\
 $|V_{e\tau}|/10^{-3}$&0.5-1.4&1.6-4.3&4.2-11.4&1.6-4.3&0.8-2.1&0.8-2.1\\
 $|V_{\tau
 e}|/10^{-4}$&2.4-10.9&7.2-32.6&4.8-21.7&4.8-21.7&1.8-8.2&4.8-21.7\\
\hline\end{tabular}\\\\
\begin{tabular}{|c|c|c|c|c|c|c|}\hline
$B_{4,5,6}$&$A_{1,3,4}$&$A_{2,5}$&$A_7^K$&$A_7^O$&$A_8^K$&$A_8^O$\\\hline
$|V_{\mu\tau}|/10^{-1}$&0.32-0.55&0.95-1.64&0.63-1.09&0.63-1.09&0.24-0.41&0.63-1.09\\
$|V_{e\mu}|/10^{-2}$&2.1-6.1&2.1-6.1&8.0-16.1&3.1-7.8&4.1-9.4&1.7-5.3\\
$|V_{e\tau}|/10^{-3}$&0.7-1.9&2.1-5.7&5.6-15.2&2.1-5.7&1.1-2.8&1.1-2.8\\
$|V_{\tau
e}|/10^{-4}$&3.2-14.5&9.6-43.5&6.4-29.0&6.4-29.0&2.4-10.9&6.4-29.0\\
\hline\end{tabular}\\\\
\begin{tabular}{|c|c|c|c|c|c|c|}\hline
$B_{7,8}$&$A_{1,3,4}$&$A_{2,5}$&$A_7^K$&$A_7^O$&$A_8^K$&$A_8^O$\\\hline
$|V_{\mu\tau}|/10^{-1}$&0.32-0.55&0.95-1.64&0.63-1.09&0.63-1.09&0.24-0.41&0.63-1.09\\
$|V_{e\mu}|/10^{-2}$&4.2-12.2&4.2-12.2&16.1-32.2&6.2-15.5&8.1-18.8&3.3-10.5\\
$|V_{e\tau}|/10^{-3}$&1.4-3.8&4.2-11.4&11.2-30.3&4.2-11.4&2.1-5.7&2.1-5.7\\
$|V_{\tau
e}|/10^{-4}$&6.4-29.0&19.3-86.9&12.8-58.0&12.8-58.0&4.8-21.7&12.8-58.0\\
\hline\end{tabular}\\\\
\begin{tabular}{|c|c|c|c|}\hline
$A_7^N$&$B_{1,2,3}$&$B_{4,5,6}$&$B_{7,8}$\\\hline $|V_{\mu
\tau}|/10^{-1}$&0.63-1.09&0.63-1.09&0.63-1.09\\
$|V_{e\mu}|/10^{-2}$&5.2-20.1&6.9-26.8&13.9-53.6\\
$|V_{e\tau}|/10^{-3}$&0&0&0\\ $|V_{\tau e}|/10^{-4}$&0&0&0\\
\hline\end{tabular}
 \end{tabular} \end{center}
\caption{Leptonic mixing angles predicted at $M_W$ by the ansatze
$A_j$ combined with any of the light ansatze $B_i$. The ranges shown
take into account the effects of varying
$\protect\alpha_S(M_Z)=0.1-0.13$, $m_b=4.1-4.4$ GeV. Changing the
Majorana mass scale $M$ does not have a significant effect when
compared to the large uncertainties inherent in the predictions.}
\label{LEPmixi}
\end{table}

\section{Filtering the Results \label{sec:filter}}
The muon and tau neutrino masses are both quite dependent upon (and
are fixed by a choice of) $M$. However, the tau neutrino mass is
relevant for cosmology, both to the cosmological bound\cite{guido}
\begin{equation}
m_{\nu_\tau} \leq \sum_{i=e,\mu,\tau} m_{\nu_i} < 100 \mbox{~eV},
\label{cosmobound}
\end{equation}
and from the point of view of structure formation in the early
universe (recent monte-carlo simulations of which predict
$m_{\nu_\tau} \sim O(5)$ eV). We shall insist that the bound in
Eq.\ref{cosmobound} is satisfied. This immediately places a lower
bound\cite{ourquadun} upon $M$ of $10^{10}$ GeV in quadruply unified
third family Yukawa coupling scenarios.

\begin{figure}
\begin{center}
\leavevmode
\hbox{\epsfxsize=5in
\epsfysize=3in
\epsffile{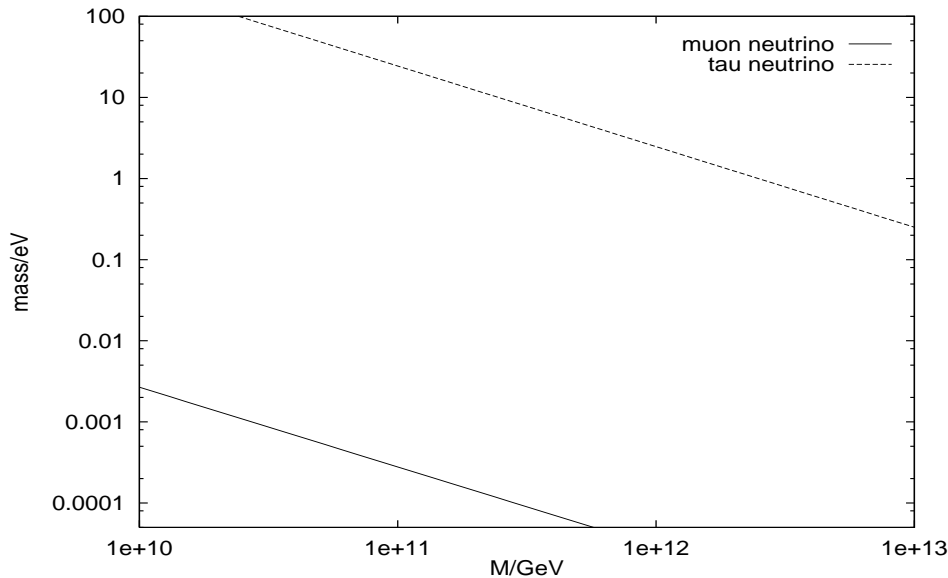}}
\end{center}
\caption{Dependence of $m_{\nu_\tau}$ and $m_{\nu_\mu}/l_N^2$ upon $M$ for
$\alpha_S(M_Z)=0.11$, $m_b=4.25$ GeV. Since the muon neutrino mass is
scaled by $1/l_N^2$, the MSW bounds upon it implied by
Eq.\ref{massMSW} must also be scaled by this number. Thus, these
bounds are dependent upon the choice of heavy ansatz $A_i$ if applied
to the figure.}
\label{fig:MSWfilt}
\end{figure}

As explained in the introduction, SUSY GUTs can provide an explanation
for the solar neutrino deficit by providing values of $\sin^2
2\theta_{e\mu}$ and $\delta m_{e\mu}^2$ that are compatible with the
MSW solution.  To yield the prediction for $m_{\nu_\mu}$, the
unification scale prediction in Eq.\ref{2famprd} implied by picking
some heavy ansatz $A_i$ is run down to low energy using the RGE in
Eq.\ref{gutsusy}. The see-saw mechanism is then employed as in
section~\ref{sec:seesaw} to obtain the mass of the light mass
eigenstate of the muon neutrino at low energies.  We shall constrain
our predictions and possible ansatze by these constraints, which to
the 95\% confidence level are
\cite{KM}
\begin{eqnarray}
\delta m_{e\mu}^2=3-10 \times 10^{-6} \mbox{~eV}^2&\Rightarrow& \delta
m_{e\mu}=1.7-3.2 \times 10^{-3} \mbox{eV} \label{massMSW} \\ sin^2
2\theta_{e\mu}= 8-150 \times 10^{-4} &\Rightarrow& |V_{e\mu}|=1.4-6.1
\times 10^{-2}. \label{mixMSW}
\end{eqnarray}
The mass constraint in Eq.\ref{massMSW} is not compatible with ansatze
$(A_7^N, B_j)$ since these predict $m_{\nu_\mu}=0$ and so we discard
these.  For any of the other particular ansatze $A_i,B_j$,
Eq.\ref{massMSW} will be true only for some restricted range of $M$,
as illustrated in Fig.\ref{fig:MSWfilt}.
This range of $M$ is valid for some range of $m_{\nu_\tau}$, as
displayed in Table~\ref{massconst}.
\begin{table}
\begin{center}
\begin{tabular}{|c|ccccc|} \hline
 & $A_{1-5}$ & $A_7^K$ & $A_7^O$ & $A_8^K$ & $A_8^O$ \\ \hline
$m_{\nu_\tau}$/eV&8-90&85-100&17-100&30-100&5-50\\ $M/10^{10}$
GeV&16-4&2-1&8-1&5-1&27-6\\ \hline
\end{tabular}
\end{center}
\label{massconst}
\caption{Constraints on the tau neutrino mass and $M$ implied by the MSW range
of $m_{e\mu}$ quoted in Eq.\protect\ref{massMSW} for each heavy ansatz
$A_i$. The upper cut-off on $m_{\protect\nu_\protect\tau}=100$ eV
implied by the cosmological bound in Eq.\protect\ref{cosmobound} has
been applied. The results are valid for
$\protect\alpha_S(M_Z)=0.1-0.13$, $m_b=4.1-4.4$ GeV.}
\end{table}
Note that in Table~\ref{massconst}, the tau neutrino mass has been cut
off at 100 eV to implement the cosmological bound. The results are
valid for any $B_i$ because the prediction of the muon neutrino mass
in Eq.\ref{2famprd} depends only upon the choice of heavy ansatz
$A_i$, as is clear from the Clebsch coefficients in
Table~\ref{ClLEPmix}.  Now we apply the MSW mixing angle filter in
Eq.\ref{mixMSW} by checking that $|V_{e\mu}|$ lies within the correct
range.  Table~\ref{LEPmixi} shows that the ansatze $(A_7^K, B_{4-8})$,
$(A_8^K, B_{7,8})$ predict a $|V_{e\mu}|$ that lies outside the
desired range and so these ansatze are discarded.

Next we constrain our results using the limits upon the $\delta m_{\mu
\tau}^2- sin^2 2\theta_{\mu \tau}$ parameter space given by the E531
experiment \cite{CHORUS}. For $\delta m_{\mu \tau}^2 > 100$ eV$^2$,
\begin{equation}
\sin^2 2 \theta_{\mu_\tau} < 5 \times 10^{-3} \Rightarrow |V_{\mu
\tau}| < 0.035.
\end{equation}
Again, the choice of light ansatz $B_j$ does not affect the prediction
of $|V_{\mu \tau}|$. The predictions of $|V_{\mu \tau}|$ displayed in
Table~\ref{LEPmixi} show that $A_{2,5}, A_7^K, A_7^O, A_8^O$ are ruled
out by this result. If one were to abandon the constraints implied by
the MSW effect, $M$ would be unconstrained and so if a larger value of
$M$ were chosen it would result in a smaller value of
$m_{\nu_\tau}$. Thus $M$ could always be chosen large enough to evade
the E531 constraints. We require successful solutions to be compatible
with the MSW range and so we do not consider these cases.

We are now left with successful ansatze $(A_{1,3,4}, B_{1-8})$,
$(A_8^K,B_{1-6})$ which simultaneously satisfy the cosmological
bounds, experimental constraints and which provide a solution to the
solar neutrino problem.

\section{Conclusions \label{sec:conc}}

We have generalised our previous operator analysis on charged fermion
masses and quark mixing angles in the SUSY 422 model \cite{patimass}
to include neutrino masses and mixing angles.  Since the model
involves no large representations of the gauge group, and has no
doublet-triplet splitting problem, the prospects for achieving string
unification in this model are very good, and some attempts in this
direction have already been made
\cite{leo2}. However here we have restricted ourselves to the
low-energy effective field theory near the scale $M_X \sim 10^{16}$
GeV, and parameterised the effects of string unification by
non-renormalisable operators whose coefficients are suppressed by
powers of $(M_X/M_S)$, where $M_S>M_X$ is some higher scale associated
with string physics.

It is worth emphasising that our previous operator analysis
\cite{patimass} uniquely specifies the Dirac sector of the neutrino
matrices without any further assumptions or parameters. This is a
simple consequence of quark-lepton unification and is also a feature
of the SO(10) model. In order to make progress we have made the simple
assumption of universal neutrino Majorana masses $M$. Since the
Majorana masses are supposed to originate from some higher-dimensional
operators in this model, as discussed in section 5, this implies some
kind of flavour symmetry in the neutrino sector.  The assumption of a
right handed neutrino Majorana mass matrix proportional to the unit
matrix in family space allows us to predict the magnitudes of all of
the leptonic CKM matrix elements in terms of quark mixing angles and
charged fermion masses, with the results being independent of $M$ to
good approximation. However the physical neutrino masses do depend on
the parameter $M$, and we have chosen to fix $M$ by requiring that the
muon neutrino has a mass in the MSW range.  Having done this the tau
neutrino mass is then predicted, as shown in Table 6. In fact the
magnitudes of all the leptonic CKM matrix elements are predicted (and
are approximately independent of $M$) as shown in Table 5.

We emphasise that the $A_i, B_i$ in
Eqs.\ref{A1}-\ref{endans},\ref{suclighti}-\ref{suclightii} were
deduced from the known charged lepton masses and quark masses and
mixing angles \cite{patimass}, and have fewer inputs than outputs in
the charged fermion sector
\cite{patimass}.
Third family Yukawa quadruple Yukawa unification (including the tau
neutrino) leads to a prediction for $m_t (\mbox{pole}) = 130-200$ GeV
and $\tan \beta = 35-65$, depending on $\alpha_S (M_Z)$ and $m_b$.
Once the MSW bound on the muon neutrino mass was imposed,
$m_{\nu_\tau}$ is predicted to be in a range relevant for both the
dark matter problem and structure formation in the early
universe. More accurate predictions could be obtained if the error on
$\alpha_S(M_Z)$ and $m_b$ were reduced.  The high values of $\tan
\beta$ required by our model (also predicted in SO(10)) can be
arranged by a suitable choice of soft SUSY breaking parameters as
discussed in ref.\cite{carloscarenayukun}, although this leads to a
moderate fine tuning problem \cite{yuki,yukiii}.  This is discussed
further in ref.\cite{patimass}.

As seen in Table 5 some of the operator combinations predict
electron-muon neutrino mixing angles outside the MSW range, and some
of them predict muon-tau mixing angles in conflict with current
experimental limits.  However some of the results are consistent with
both, as summarised in Table~\ref{summary}.  These results are
associated with the Clebsch relations in Eqs.
\ref{2ndpred},\ref{1stpred},\ref{2famprd},\ref{mutaupred}
which permits a qualitative understanding of the numerical results.
For example for $A_{1,3,4}$ ($A_8^K$) we have $\left(
\frac{\lambda_\mu}{\lambda_s} \right)_{M_{X}}=3(4)$, $\left(
\frac{\lambda_{\nu_\mu}}{\lambda_{c}} \right)_{M_X}=3(3/2)$ and
$\left( \frac{|V_{\mu \tau}|}{|V_{cb}|}\right) _{M_X}=1(3/4)$.  Thus
once the muon neutrino mass has been adjusted into the correct MSW
range, the Clebsch relations tell us that the tau neutrino mass
predicted by $A_8^K$ will be about four times larger than that
predicted by $A_{1,3,4}$. Similarly, the level of muon-tau neutrino
mixing is seen to be slightly larger in the case of $A_{1,3,4}$ than
for $A_8^K$, due to its slightly larger Clebsch. In fact the values of
$\delta m_{\tau \mu}^2$ and $\sin^2 2\theta_{\mu
\tau}$ quoted in
Table~\ref{summary} are on the edge of the present exclusion zone.  If
our scheme is correct, muon-tau neutrino oscillations will be seen
very soon by the CHORUS and NOMAD experiments.

\begin{table}
\begin{center}
\begin{tabular}{|c|c|c|} \hline
&$A_{1,3,4}$ & $A_8^K$ \\ \hline & $\delta m_{\mu \tau}^2=64-8100$
 eV$^2$ & $\delta m_{\mu \tau}^2=900-10^{4}$ eV$^2$ \\ & $\sin^2
 2\theta_{\mu \tau}=4.1-5.0 \times 10^{-3}$ & $\sin^2 2\theta_{\mu
 \tau}=2.3-5.0 \times 10^{-3}$ \\ \hline &
 $\sin^22\theta_{e\mu}=1.0-2.1 \times 10^{-3}$ &
 $\sin^22\theta_{e\mu}=3.6-20.0 \times 10^{-3}$ \\ $B_{1,2,3}$ &
 $\sin^22\theta_{e\tau}=1.1-8.1\times 10^{-6}$ &
 $\sin^22\theta_{e\tau}=2.5-18.1\times 10^{-6}$ \\ &
 $\sin^22\theta_{\tau e}=2.3-47.2\times 10^{-7}$ &
 $\sin^22\theta_{\tau e}=1.3-26.6\times 10^{-7}$ \\ \hline &
 $\sin^22\theta_{e\mu}=1.8-14.8 \times 10^{-3}$ &
 $\sin^22\theta_{e\mu}=6.7-35.9 \times 10^{-3}$ \\ $B_{4,5,6}$ &
 $\sin^22\theta_{e\tau}=2.0-14.4\times 10^{-6}$ &
 $\sin^22\theta_{e\tau}=4.4-32.2\times 10^{-6}$ \\ &
 $\sin^22\theta_{\tau e}=4.1-84.0\times 10^{-7}$ &
 $\sin^22\theta_{\tau e}=2.3-47.3\times 10^{-7}$ \\ \hline
 &$\sin^22\theta_{e\mu}=1.0-2.1 \times 10^{-3}$ & \\ $B_{7,8}$ &
 $\sin^22\theta_{e\tau}=7.8-57.4\times 10^{-6}$ & Not in MSW range.\\
 & $\sin^22\theta_{\tau e}=16.5-336\times 10^{-7}$ & \\ \hline
\end{tabular}
\end{center}
\caption{A summary of our predictions which satisfy the
MSW solar neutrino mass and mixing data, the cosmological tau neutrino
mass upper bound and the E531 tau-muon neutrino mixing exclusion
zone. The tau neutrino mass prediction depends on the parameter $M$
which has been chosen so that the muon neutrino mass lies in the MSW
range.  The electron neutrino mass is effectively zero.  The mixing
angle predictions are approximately independent of $M$, although they
do rely on the assumption of universality.  Note that the
$\protect\delta m_{\protect\mu \protect\tau}^2$ and $\protect\sin^2
2\protect\theta_{\protect\mu
\protect\tau}$ predictions are valid for each of the light ansatze $B_i$.}
\label{summary}
\end{table}

It is clear from Table~\ref{summary} that $A_8^K$ implies tau neutrino
masses which are too high compared to the $\sim 5$ eV masses suggested
by recent calculations on structure formation in the early
universe. However, ansatze $A_{1,3,4}$ predict the tau neutrino mass
to be as low as about 10 eV, a number in the right ball park for
structure formation, so on this basis one may wish to discard the
$A_8^K$ ansatze. Restricting our attention to $A_{1,3,4}$ only, we see
from Table~\ref{kvals} that $B_{4}$ predicts $\left(
\frac{\lambda_d}{\lambda_e} \right)_{M_{X}}=3$ (the GJ prediction)
while $B_{1,5}$ predicts $\left( \frac{\lambda_d}{\lambda_e}
\right)_{M_{X}}=4$.  The $B_6$,$B_7$ are excluded by the down quark
mass, and $B_3$,$B_8$ give rather small down masses while $B_2$ gives
a rather large down mass and these may soon be excluded by more
accurate measurements of the down quark mass and $\alpha_3(M_Z)$
\cite{patimass}.  If one discounts these cases, then the list of
possibilities is reduced to $A_{1,3,4}\otimes B_{1,4,5}$.

Table~\ref{summary} shows that the combination of $A_{1,3,4}$ with
$B_{1}$ involves $\sin^22\theta_{e\mu}\approx 1-2 \times 10^{-3}$,
while $A_{1,3,4}$ with $B_{4,5}$ yields $\sin^22\theta_{e\mu}\approx
2-15 \times 10^{-3}$.  These two possibilities may be therefore
distinguished by data from future solar neutrino experiments, assuming
the MSW mechanism to be operative.  Since $B_{4,5}$ give different
down quark mass predictions, the combination of improved down quark
mass determinations and solar neutrino data may eventually allow the
particular ansatz $B_i$ which is combined with $A_{1,3,4}$ to be
uniquely specified. The individual $A_{1,3,4}$ cannot be
differentiated by such a bottom-up procedure, however.

To summarise, in this paper we have exploited our recent operator
analysis of the charged fermion sector \cite{patimass} which fixes the
Dirac part of the neutrino mass matrix, and supplemented it by the
simple assumption of universal Majorana neutrino masses $M$. We have
shown how the MSW solar neutrino data, plus the cosmological bound on
the tau neutrino mass and the present limit on muon-tau neutrino
mixing reduces a large number of possible ansatze to just a few.  We
find the idea of predicting all the neutrino masses and mixing angles
in terms of the known fermion spectrum plus one additional parameter
$M$, essentially a generalisation of the back-of-the-envelope estimate
in Eq.\ref{envelope}, to be a very attractive and simplifying idea.
Within the framework of a particular model, we have shown how this
idea of universal Majorana masses has strong predictive power in the
neutrino sector, with imminently testable cosmological, astrophysical
and terrestrial consequences.

\section*{Acknowledgements}
B.C.Allanach would like to thank PPARC for financial support in the
duration of this work.

\newpage

\section*{Appendix}

In our particular case $A_1, B_1$ the Clebsch coefficients in
Eq.\ref{firstyuk} are given by
\begin{equation} \begin{array}{cccc}
x_{12}^U =0 & x_{12}^D = 1 & x_{12}^E = 1 & x_{12}^N=0\\ x_{21}^U =1 &
x_{21}^D = 3 & x_{21}^E = 9/4& x_{21}^N=3/4\\ x_{22}^U =1 & x_{22}^D =
-1 & x_{22}^E = 3 & x_{22}^N=-3\\ {x'}_{22}^U = 1 & {x'}_{22}^D = 1 &
{x'}_{22}^E = -3 &{x'}_{22}^N=-3 \\ x_{32}^U = 1 & x_{32}^D = -1 &
x_{32}^E = -1&x_{32}^N=1 \\
\end{array}\label{clebsch2}
\end{equation}
and ${x'}_{12}^U \neq 0$.

At $M_X$, we have the freedom to rotate the phases of $F^i$ and
$\bar{F}_j$, since this leaves the lagrangian of the high energy
theory invariant. In doing this we rotate away 5 phases in the
matrices since there are only 5 {\em relative} \/phases:
\begin{eqnarray}
\left[\begin{array}{c}
\bar{F}_1 \\ \bar{F}_2 \\ \bar{F}_3 \end{array}\right] &\rightarrow&
\left[\begin{array}{ccc}
e^{-i (\phi_{32} - \phi_{12})} & 0 & 0 \\ 0 & e^{-i (\phi_{32} -
\phi_{22}^D)} & 0 \\ 0 & 0 & 1 \\ \end{array}\right]
\left[\begin{array}{c}
\bar{F}_1 \\ \bar{F}_2 \\ \bar{F}_3 \end{array}\right]
\nonumber \\
\left[\begin{array}{c}
F_1 \\ F_2 \\ F_3 \end{array}\right] &\rightarrow&
\left[\begin{array}{ccc}
e^{-i (-\phi_{32}+\phi_{22}^D - \phi_{21})} & 0 & 0 \\ 0 &e^{i
\phi_{32}} & 0 \\ 0 & 0 & e^{i \phi_{33}} \\ \end{array}\right]
\left[\begin{array}{c}
F_1 \\ F_2 \\ F_3 \end{array}\right].
\label{phaserots}
\end{eqnarray}
Note that the first phase rotation performed in Eq.\ref{phaserots}
introduces phases into the Majorana mass matrix $M\sim
\mbox{diag}(e^{2i (
\phi_{32}-\phi_{12})}, e^{2i(\phi_{32}-\phi_{22}^D)},1)$.

Once the Yukawa matrices have been derived from the phase rotations in
Eq.\ref{phaserots}, $Y^{U,N}$ are made real by multiplying by the
phase matrices
\begin{eqnarray}
Y^U &\rightarrow&
\left[\begin{array}{ccc}
e^{-i \bar{\phi}_{12}} & 0 & 0 \\ 0 & e^{-i \bar{\phi}_{22}} & 0 \\ 0
& 0 & 1 \\
\end{array}\right]
Y^{U} \left[\begin{array}{ccc} e^{i \bar{\phi}_{22}^U} & 0 & 0 \\ 0 &
1 & 0 \\ 0 & 0 & 1 \\
\end{array}\right] \nonumber \\
Y^N &\rightarrow&
\left[\begin{array}{ccc}
1 & 0 & 0 \\ 0 & e^{-i \bar{\phi}_{22}} & 0 \\ 0 & 0 & 1 \\
\end{array}\right]
Y^{N} \left[\begin{array}{ccc} e^{i \bar{\phi}_{22}^U} & 0 & 0 \\ 0 &
1 & 0 \\ 0 & 0 & 1 \\
\end{array}\right]
,\label{phasegone}
\end{eqnarray}
where we have defined $\bar{\phi}_{22} \equiv \phi_{22}^U -
\phi_{22}^D$ and $\bar{\phi}_{12}  \equiv \phi_{12}' - \phi_{12}$.
This amounts to a phase redefinition of the $(U_R)^c$, $U_L$, $\nu_R$
and $\nu_L$ fields.

To diagonalise the real matrices obtained from the above phase
rotations, we first diagonalise the heavy 2 by 2 submatrices, then the
light submatrices as shown below,
\begin{eqnarray}
Y^D \rightarrow
\left[\begin{array}{ccc}
 \tilde{c}_2 & \tilde{s}_2 & 0 \\ - \tilde{s}_2 & \tilde{c}_2 & 0 \\ 0
& 0 & 1 \end{array}\right]
\left[\begin{array}{ccc}
 1& 0 & 0 \\ 0 & \tilde{c}_4 & \tilde{s}_4 \\ 0 & -\tilde{s}_4 &
\tilde{c}_4 \\ \end{array}\right] & Y^D &
\left[\begin{array}{ccc}
1 & 0 & 0 \\ 0 & \bar{c}_4 & -\bar{s}_4 \\ 0 & \bar{s}_4 & \bar{c}_4
\\ \end{array}\right]
\left[\begin{array}{ccc}
c_2 & -s_2 & 0 \\ s_2 & c_2 & 0 \\ 0 & 0 & 1 \\ \end{array}\right]
\nonumber \\ Y^U \rightarrow
\left[\begin{array}{ccc}
 \tilde{c}_1 & \tilde{s}_1 & 0\\ -\tilde{s}_1 & \tilde{c}_1 & 0\\ 0 &
 0 & 1 \end{array}\right]
\left[\begin{array}{ccc}
1 & 0 & 0 \\ 0 & \tilde{c}_3 & \tilde{s}_3 \\ 0 & -\tilde{s}_3 &
\tilde{c}_3 \\ \end{array}\right] & Y^U &
\left[\begin{array}{ccc}
1 & 0 & 0 \\ 0 & \bar{c}_3 & -\bar{s}_3 \\ 0 & \bar{s}_3 & \bar{c}_3
\\ \end{array}\right]
\left[\begin{array}{ccc}
c_1 & -s_1 & 0 \\ s_1 & c_1 & 0 \\ 0 & 0 & 1 \\ \end{array}\right]
\nonumber \\
Y^E \rightarrow
\left[\begin{array}{ccc}
 \tilde{c}^l_2 & \tilde{s}^l_2 & 0 \\ - \tilde{s}^l_2 & \tilde{c}^l_2
& 0 \\ 0 & 0 & 1 \end{array}\right]
\left[\begin{array}{ccc}
 1& 0 & 0 \\ 0 & \tilde{c}^l_4 & \tilde{s}^l_4 \\ 0 & -\tilde{s}^l_4 &
\tilde{c}^l_4 \\ \end{array}\right] & Y^E &
\left[\begin{array}{ccc}
1 & 0 & 0 \\ 0 & \bar{c}^l_4 & -\bar{s}^l_4 \\ 0 & \bar{s}^l_4 &
\bar{c}^l_4 \\ \end{array}\right]
\left[\begin{array}{ccc}
c_2^l & -s_2^l & 0 \\ s_2^l & c_2^l & 0 \\ 0 & 0 & 1 \\
\end{array}\right] \nonumber \\ Y^N \rightarrow
\left[\begin{array}{ccc}
 \tilde{c}^l_1 & \tilde{s}^l_1 & 0\\ -\tilde{s}^l_1 & \tilde{c}^l_1 &
 0\\ 0 & 0 & 1 \end{array}\right]
\left[\begin{array}{ccc}
1 & 0 & 0 \\ 0 & \tilde{c}^l_3 & \tilde{s}^l_3 \\ 0 & -\tilde{s}^l_3 &
\tilde{c}^l_3 \\ \end{array}\right] & Y^N &
\left[\begin{array}{ccc}
1 & 0 & 0 \\ 0 & \bar{c}^l_3 & -\bar{s}^l_3 \\ 0 & \bar{s}^l_3 &
\bar{c}^l_3 \\ \end{array}\right]
\left[\begin{array}{ccc}
c_1^l & -s_1^l & 0 \\ s_1^l & c_1^l & 0 \\ 0 & 0 & 1 \\
\end{array}\right], \nonumber \\ \label{diags}
\end{eqnarray}
where $c_i^{(l)}, \bar{s}_i^{(l)}$ refer to $\cos \theta_i^{(l)}$ and
$\sin
\bar{\theta}_i^{(l)}$ respectively. Note that since $Y^{U,D,E,N}$ are not
symmetric $\tilde{c}_i^{(l)}, \tilde{s}_i^{(l)}$ are independent of
$c_i^{(l)},
\bar{s}_i^{(l)}$.

The quark mixing angles are given by $\bar{s}_4=-x_{32}^DH_{32}/
H_{33}$, $s_2=-x_{21}^DH_{21}/(\lambda_s)_{M_X}$,
$s_1=-x_{21}^UH_{21}/(\lambda_c)_{M_X}$ and
$\bar{s}_3=-x^U_{32}H_{32}/H_{33}$.  Note that in the limit ${O_{12}}'
\rightarrow 0$ the up quark is massless in the model because it is
generated by the small operator $(\lambda_u)_{M_X}=-{x'}_{12}^U
x_{21}^UH_{12}' H_{21} / (\lambda_c)_{M_X}$.  The lepton mixing angles
are $\bar{s}^l_4=-x_{32}^EH_{32}/ H_{33}$,
$s_2^l=-x_{21}^EH_{21}/(\lambda_\mu)_{M_X}$,
$s_1^l=-x_{21}^NH_{21}/(\lambda_{\nu_\mu})_{M_X}$ and
$\bar{s}^l_3=-x^N_{32}H_{32}/H_{33}$.  We also denote
$\theta_3^{(l)}=\bar{\theta}_3^{(l)}-\bar{\theta}_4^{(l)}$ and $\phi =
-
\bar{\phi}_{22}$.

Noting that $|V_{\mu \tau}| \sim s_3^l = (x_{32}^N-x_{32}^E)
H_{32}/H_{33}$ and that $|V_{cb}| \sim s_3 = (x_{32}^U-x_{32}^D)
H_{32}/H_{33}$ we make the prediction in Eq.\ref{mutaupred} where $m =
(x_{32}^N-x_{32}^E)/(x_{32}^U-x_{32}^D)$ in this case.  We also have
that $|V_{e \tau}| \sim s_1^l s_3^l \Rightarrow |V_{e
\tau}|/|V_{\mu \tau}| = s_1^l = -x_{21}^N H_{21}/ \lambda_{\nu_\mu}$.
{}From $V_{CKM}$, $H_{21}=- |\lambda_c| |V_{ub}| / ( x_{21}^u
|V_{cb}|)$. Substituting Eqs.\ref{mutaupred},\ref{2famprd} yields
Eq.\ref{etauub} where $n= m x_{21}^N / (x_{21}^Ul_N)$. When $H_{21}$
is substituted into the ratio $|V_{\tau e}/V_{\mu \tau}| \sim
s_2^l=x_{21}^E H_{21}/ \lambda_\mu$, we obtain our next prediction as
in Eq.\ref{taue}. Our final prediction is for $|V_{e \mu}| \sim
|-s_2^l e^{i \phi} + s_1^l|$ which is made in terms of the phase
$\phi$. As above, we substitute the angles $\theta_1^l, \theta_2^l$ to
obtain Eq.\ref{emupred} where $p=x_{21}^N/(l_N x_{21}^U), q=l_N
x_{21}^E / x_{21}^N$.

 \end{document}